\documentclass[9pt]{article}

\usepackage[latin9]{inputenc}
\usepackage[T1]{fontenc}
\usepackage{longtable}
\usepackage[a4paper, total={19cm, 24.7cm}, left=1.3cm]{geometry}
\usepackage{float}
\usepackage{graphicx}
\usepackage{epstopdf}
\usepackage{subfigure}
\usepackage{amssymb}
\usepackage{fancybox}
\usepackage{amsmath}
\usepackage{epsfig}
\usepackage{dsfont}
\usepackage{fancyhdr}
\usepackage{picins}

\usepackage{multicol}
\usepackage{color}
\usepackage{array}
\usepackage{multirow}
\usepackage{dcolumn}
\usepackage[square,numbers,sort&compress]{natbib}

\renewcommand{\section}[2]{} 
\begin{document}
\begin{Huge}
\noindent {\bf Direct detection of the $^{229}$Th nuclear}\\[0.3cm] 
\noindent {\bf clock transition}\\[0.5cm]
\end{Huge}
Lars von der Wense$^{1}$,\let\thefootnote\relax\footnote{$^{1}$Ludwig-Maximilians-University Munich, 85748 Garching, Germany. $^{2}$ GSI Helmholtzzentrum für Schwerionenforschung GmbH, 64291 Darmstadt, Germany. $^{3}$ Helmholtz Institute Mainz, 55099 Mainz, Germany. $^{4}$ Johannes Gutenberg University, 55099 Mainz, Germany.} Benedict Seiferle$^{1}$, Mustapha Laatiaoui$^{2,3}$, Jürgen B. Neumayr$^{1}$, Hans-Jörg Maier$^{1}$, Hans-Friedrich Wirth$^{1}$, Christoph Mokry$^{3,4}$, Jörg Runke$^{2,4}$, Klaus Eberhardt$^{3,4}$, Christoph E. Düllmann$^{2,3,4}$, Norbert G. Trautmann$^{4}$ \& \\ Peter G. Thirolf$^{1}$\\[0.2cm]
{\bf 
Today's most precise time and frequency measurements are performed with optical atomic clocks. However, it has been proposed that they could potentially be outperformed by a nuclear clock, which employs a nuclear transition instead of the atomic shell transitions used so far. By today there is only one nuclear state known which could serve for a nuclear clock using currently available technology, which is the isomeric first excited state in $^{229}$Th. Here we report the direct detection of this nuclear state, which is a further confirmation of the isomer's existence and lays the foundation for precise studies of the isomer's decay parameters. Based on this direct detection the isomeric energy is constrained to lie between 6.3 and 18.3~eV, and the half-life is found to be longer than 60~s for $^{229\mathrm{m}}$Th$^{2+}$. More precise determinations appear in reach and will pave the way for the development of a nuclear frequency standard.}
\begin{multicols*}{2}
\vspace{-1cm}
\noindent
The first excited nuclear state in $^{229}$Th is one of the most exotic states in the whole nuclear landscape, as among all presently known 176,000 nuclear levels \cite{gov}, it possesses the lowest excitation energy of only about $7.8$~eV \cite{Beck1,Beck2}. While there is just one other nuclear excitation known to have a transition energy below 1~keV \cite{gov} ($^{235\mathrm{m}}$U, 76~eV), typical nuclear excitation energies are $10^4$ to $10^6$ times larger \cite{wallet} (Fig.~1).\\
The $^{229}$Th nucleus was first considered in 1976 by Kroger and Reich to possess an isomer (a metastable nuclear state) with excitation energy below 100~eV \cite{Kroger_Reich}. Further measurements supported its existence \cite{Burke_Garrett,Burke2} and led to a stepwise improvement to $(3.5\pm1.0)$~eV excitation energy in 1994 \cite{Helmer_Reich2}. However, in 2007 a microcalorimetric measurement suggested a value of $(7.8\pm0.5)$~eV, corresponding to a wavelength near 160~nm for radiation emitted in the decay to the ground state \cite{Beck1,Beck2}.\\ 
This uniquely low nuclear transition energy can potentially bridge the fields of nuclear and atomic physics, as it conceptually allows for optical laser excitation of a nuclear transition \cite{Tkalya1}. This in turn has stimulated thoughts about transferring existing knowledge of laser manipulation of the electronic shell to a nuclear system, leading to highly interesting applications such as a nuclear laser \cite{Tkalya3}, nuclear quantum optics \cite{Buervenich}, and a nuclear clock \cite{Peik1}.\\
Besides the low excitation energy, a radiative isomeric half-life in the range of minutes to hours has been predicted \cite{Ruchowska,Karpeshin1,Tkalya4}, resulting in a relative linewidth as low as $\Delta E/E\approx 10^{-20}$. These unique features render this transition an ideal candidate for a nuclear clock \cite{Peik1}, which may outperform existing atomic-clock technology due to potentially improved compactness and expectedly higher resilience against external influences \cite{Peik2}. Two ways to establish a nuclear clock are currently being investigated; one based on $^{229}$Th$^{3+}$ stored in a Paul trap \cite{Zimmermann,Campbell1,Campbell2}, the other one based on $^{229}$Th embedded in a crystal-lattice environment \cite{Peik1,Tkalya2,Jeet,Kazakov1,Simon}.\\
The immediate impact and far reaching implications of a nuclear clock become clear when considering current applications of existing atomic-clock technology \cite{Nicholson}. Moreover, a nuclear clock promises intriguing applications in fundamental physics, e.g., the investigation of possible time variations of fundamental constants \cite{Flambaum1,Hayes,Litvinova,Rellergert}.
\begin{figure}[H]
 \begin{center}
 \includegraphics[width=8.9cm]{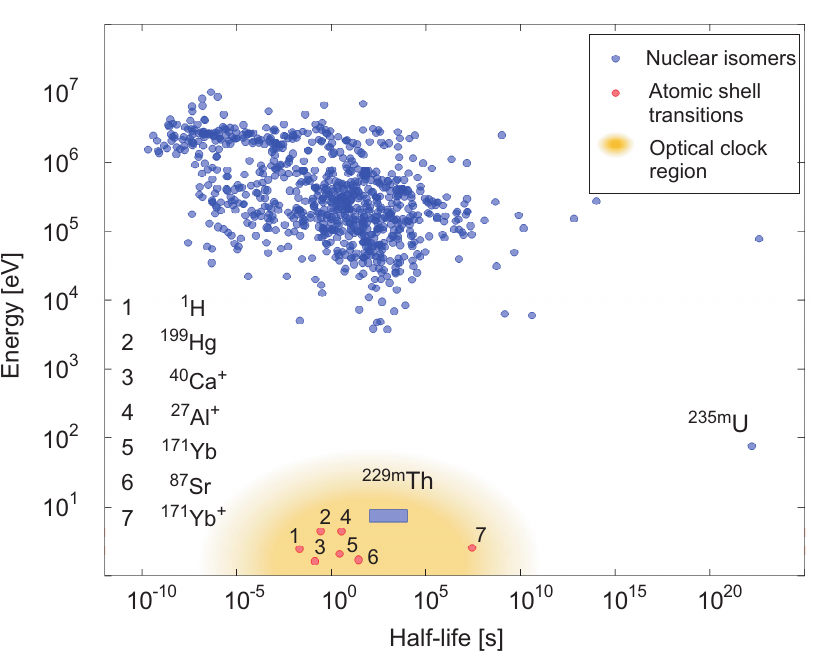}
 \vspace{-0.0cm}
  \caption{\small {\bf Energy-half-life distribution of known nuclear isomeric states \cite{wallet} (blue) and selected atomic shell transitions used for frequency metrology (red, numbered).} The orange region highlights the parameter space currently accessible for optical clocks. $^{229\mathrm{m}}$Th (expected region shown as box) exhibits a uniquely low excitation energy and is the only known promising isomer for the development of a nuclear-based frequency standard using existing technology. One other nuclear isomer with an energy below $10^3$~eV is known ($^{235\mathrm{m}}$U), however, at a significantly longer half-life. Purely radiative half-lives are shown for $^{229\mathrm{m}}$Th and $^{235\mathrm{m}}$U, this being the relevant parameter for the development of a nuclear clock.}
 \label{energy_halflife}
 \end{center}
\end{figure}
\end{multicols*}
\begin{figure}[H]
 \begin{center}
 \vspace{-1.5cm}
 \includegraphics[width=18.3cm]{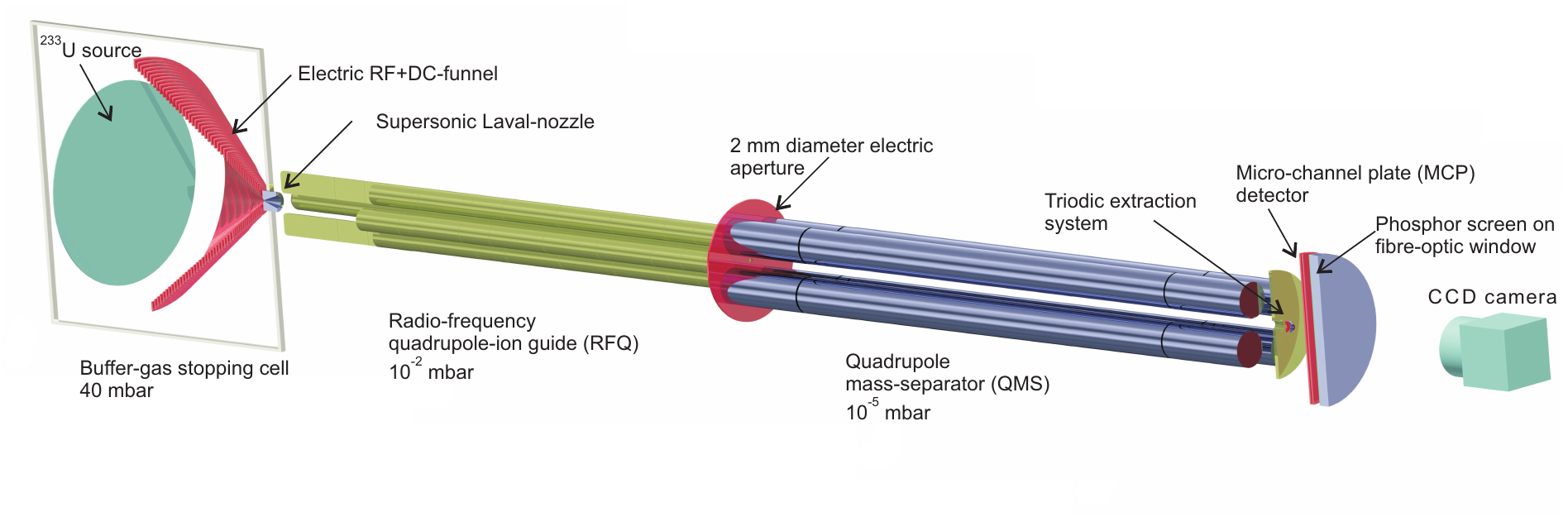}
 \vspace{-1cm}
  \caption{\small {\bf Schematic of the experimental setup.} The $^{233}$U source is mounted in front of an RF+DC-funnel placed in a buffer-gas stopping cell \cite{Neumayr1}. $^{229}$Th $\alpha$-recoil ions, emitted from the source, are extracted for ion-beam production in a radio-frequency quadrupole system (RFQ). After mass purification of the ion beam with the help of a quadrupole mass-separator (QMS), the ions are attracted at low kinetic energy (for soft landing) onto the surface of a micro-channel plate detector (MCP). There the $^{229}$Th isomeric decay signals are detected (for details see text and Methods).}
 \label{setup}
 \end{center}
\end{figure}
\vspace{-1cm}
\begin{multicols*}{2}
\noindent To date, experimental knowledge of the isomer has been inferred indirectly \cite{Beck1,Beck2,Kroger_Reich,Burke_Garrett,Burke2,Helmer_Reich2}. However, a direct detection was still pending. Such a direct detection would not only give further evidence for the isomer's existence, but also pave the way for precise studies of the half-life, excitation energy and decay mechanism of the isomeric state, which are the basis for a direct optical excitation \cite{Matinyan}. This has motivated significant experimental effort aimed at further validation of the isomer's existence \cite{Raeder,Sonnenschein,Kazakov2} and direct detection of the isomeric deexcitation \cite{Jeet,Irwin,Richardson,Utter,Zhao,Peik4,Yamaguchi}. For a detailed overview we refer the reader to the recent review of Peik and Okhapkin and references therein \cite{Peik3}. Despite decade-long efforts, none of these previous attempts has conclusively reported the isomer's direct detection. We report here on the direct observation of this elusive isomeric decay. This direct detection paves the way for the precise determination of all decay parameters relevant for optical excitation.\\[0.5cm]
\begin{large}
\noindent {\bf \textcolor{blue}{Experimental setup}}\\[0.2cm]
\end{large}
Decay of the $^{229}$Th isomeric state in the neutral thorium atom occurs predominantly by internal conversion (IC) under emission of an electron \cite{Karpeshin1,Tkalya4}, which is used as a key signature for identifying the $^{229}$Th isomer (3/2$^+$[631]) to ground state (5/2$^+$[633]) deexcitation. A short half-life in the $\mu$s range was predicted for this case \cite{Karpeshin1,Tkalya4}. This is because the 6.31 eV first ionisation potential of thorium is below the suggested energy of the isomeric transition. In a higher charge state (i.e. thorium ions), the IC process is energetically forbidden and radiative decay may dominate. In this case, the half-life is expected to increase significantly to minutes or hours. While searches for an IC decay with a half-life of a few ms or longer for neutral thorium have already been conducted \cite{Swanberg}, our experimental setup \cite{Wense1}, as shown in Fig.~2, was designed for the detection of a low-energy IC decay of shorter half-life. A schematic of the experimental process is shown in Extended Data Fig.~1.\\
The isomeric state in $^{229}$Th can be populated via a 2\% decay branch in the $\alpha$ decay of $^{233}$U \cite{Barci}. For detection of the isomer, a $^{233}$U source is placed in a buffer-gas stopping cell \cite{Neumayr1} (Extended Data Fig.~2) into which $^{229}$Th ions, produced in the $\alpha$ decay of $^{233}$U, are recoiling, along with $^{229}$Th daughter products if present in the source. These $\alpha$-recoil ions are stopped in 40 mbar of ultra-pure helium. Removing the up to 84~keV kinetic recoil-energy (significantly greater than the few eV isomer energy) is essential for the experiment. During the stopping process charge exchange occurs producing predominantly thorium in the 2+ and 3+ charge states. These ions are guided by an electric field through a radio-frequency (RF) and direct-current (DC) funnel system towards the buffer-gas stopping cell exit, where they are extracted by a supersonic Laval-nozzle and injected into a radio-frequency quadrupole (RFQ) structure. While the ions are guided by the electric fields provided by the RFQ, the remaining ambient helium gas pressure leads to phase-space cooling, such that a recoil-ion beam with sub-mm diameter is formed at the RFQ exit. There, most of the daughter nuclides from the $^{233}$U decay chain are still present, some of which are short-lived $\alpha$ or $\beta^-$ emitters. A quadrupole mass-separator (QMS) is used for ion-beam purification, such that only $^{229}$Th remains. Subsequently, the thorium ions are guided with the help of a triodic guidance structure with a 2-mm diameter orifice towards a micro-channel plate (MCP) detector, used for low-energy electron detection. The ions are collected in soft landing at low kinetic energy (50-75~eV, depending on the charge state) directly on the MCP detector (operated at $-25$~V surface voltage), which is placed in front of a phosphor screen. The latter is monitored by a charge-coupled device (CCD) camera, allowing for a spatially resolved signal detection.\\[0.5cm]
\begin{large}
\noindent{\bf \textcolor{blue}{Isomer detection}}\\[0.2cm]
\end{large}
Because stopping and extraction of $^{229}$Th occurs in the form of ions and takes only a few ms, there is no significant isomer decay during time of flight. However, when the ions come into contact with the MCP surface, charge exchange occurs forming neutral thorium atoms for which the rapid IC expectedly dominates the decay of the isomeric state. This process releases a conversion electron, which is accelerated into a microchannel of the MCP detector, triggering the emission of secondary electrons. The electron 'cloud' thusly produced is accelerated towards the phosphor screen, where the electronic-impact signal is converted into visible light that is detected with the CCD camera. This detection technique reveals a significant similarity to the MCP-based detection of metastable molecular states in chemistry \cite{Jongma} and has already been successfully applied to the detection of
\begin{figure}[H]
 \begin{center}
 \includegraphics[width=8.9cm]{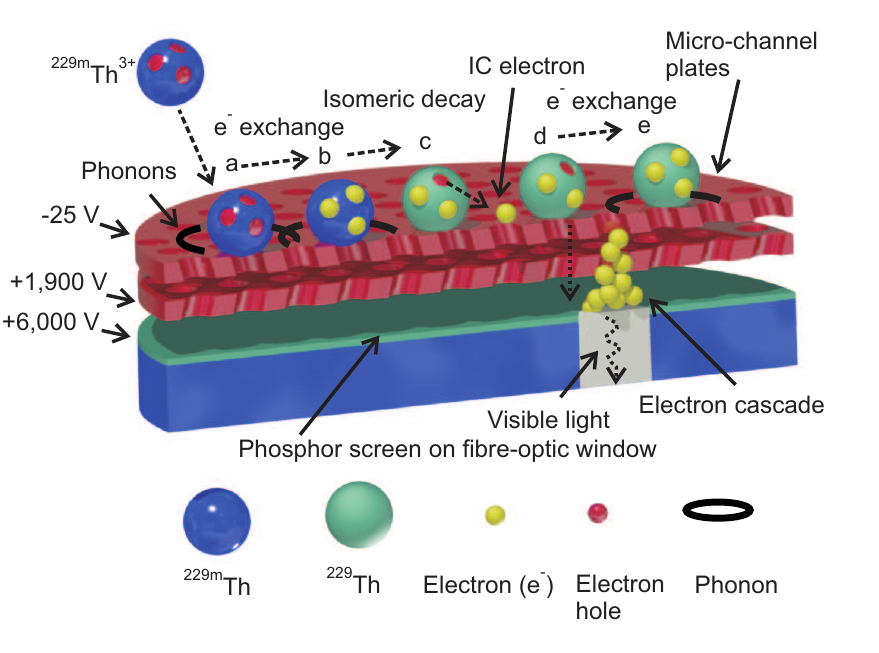}
 \caption{\small {\bf Schematic drawing of the isomer detection process.} a: A $^{229\mathrm{m}}$Th$^{3+}$ ion impinges on the MCP surface. The thorium ion in the isomeric state is visualised as a blue sphere. b: Electron capture on the surface. The energy is dissipated in form of phonons (indicated as black circles). Electrons are visualised as yellow spheres. c: An IC electron is released by the isomeric decay. d: The IC electron triggers a secondary-electron cascade, which is accelerated towards the phosphor screen. e: The hole, left by the IC process, is closed by electron capture on the MCP surface. Again phonons are produced.}
 \label{concept_draw}
 \end{center}
\end{figure}
\vspace{-0.5cm}
\noindent $^{235\mathrm{m}}$U \cite{Swanberg}. A schematic drawing of the detection process on the MCP surface from a microscopic perspective is shown in Figure~3 (see also Methods section).\\
The $^{233}$U source (in the following denoted as source 1) consists of a layer of $^{233}$UF$_4$ (of activity level $\sim200$~kBq) that was evaporated onto a 20-mm diameter stainless steel plate. A complete mass scan of ions extracted from this source is shown in Fig.~4~a. We measured the $^{229}$Th$^{3+}$ ion extraction rate from source 1 to be about $10^{3}$ s$^{-1}$ \cite{Wense2}. Assuming that 2\% of the ions are in the isomeric state \cite{Barci} and also accounting for an MCP detection efficiency for low energy electrons of about 1.5\% \cite{Goruganthu}, a count rate of $\sim0.3$~cts/s is expected. The isomeric-decay signal as obtained when extracting $^{229}$Th$^{3+}$ for 2,000~s is shown in Fig.~4~c. Signals were acquired within a centered field of view as obtained within a 20-mm diameter aperture (see Methods for details of image readout). The spatially integrated decay count rate is ($0.25\pm0.10$)~cts/s and in good agreement with the expectations. The error was estimated to also account for changes in the $^{229}$Th$^{3+}$ extraction efficiency.
The MCP exhibits a low dark count rate of 0.01~cts/s mm$^2$, leading to a signal to background ratio of about 8:1. An overview over different measurements performed under the same conditions is shown in Fig.~4~b. Each row corresponds to an individual uranium source, as will be detailed in the following section, while each column corresponds to a different extracted ion species, as indicated by the arrows from the mass scan. Clear signals are seen when extracting $^{229}$Th$^{2+}$ and $^{229}$Th$^{3+}$, respectively (Fig.~4~b, first row). For completeness, measurements were also performed while extracting $^{229}$Th$^{1+}$. In this case, no signal could be obtained, which might be attributed to the very low extraction efficiency of just 0.3\% for Th$^{1+}$, compared to 5.5\% for Th$^{2+}$ and 10\% for Th$^{3+}$ \cite{Wense2}.\\[0.5cm] 

\begin{large}
\noindent{\bf \textcolor{blue}{Signal identification}}\\[0.2cm]
\end{large}
In order to prove that the detected signal originates from the $^{229}$Th isomeric decay, comparative measurements were performed which allowed us to exclude all potential background sources. These can be grouped into four categories: (A) background attributed to the kinetic energy or charge state of the impinging ions, (B) background signals from setup components ($^{233}$U source, buffer-gas stopping and extraction, QMS, MCP detection system), (C) signals originating from the thorium atomic shell (long-lived excited states or chemical reactions on the MCP surface) and (D) signals caused by short-lived nuclides or other isomers (not of $^{229}$Th). Most of the possible background effects were excluded in several ways. An overview is shown in the Extended Data Table~1, to which the given measurement numbers refer.\\
Ionic energy, as carried in form of the momentum or ion charge state, may lead to the release of electrons on the MCP surface. In order to exclude this type of background (type~A), a $^{233}$U$^{2+}$ mass peak (originating from sputtering of the source), which has a similar intensity as the $^{229}$Th$^{2+}$ mass peak (Fig.~4~a), is used for comparison (Extended Data Table~1 no.~1). Within 2,000~s of continuous extraction of $^{233}$U$^{2+}$ no MCP signal was obtained (Fig.~4~b, first row).\\
Furthermore, a measurement of the signal intensity as a function of the MCP surface voltage was carried out for $^{229}$Th$^{2+}$ and $^{233}$U$^{2+}$ (Extended Data Table~1 no.~2, Fig.~5 a). For this purpose, each isotope was extracted 1,200 s for every data point. For MCP surface voltages between $-100$ V and $-40$ V the remaining ion-impact signal decreases as the kinetic energy of the ions is reduced. While the uranium signal is effectively reduced to zero, a thorium signal remains. A sharp cut-off of this signal occurs at zero kinetic energy, when the ions can no longer approach the MCP surface. An enhancement of the signal intensity is observed close before the cut-off, attributed to IC electrons back-attracted into the MCP surface. The absence of a similar sharp cut-off for uranium clearly excludes any cause of the signal by ion impact or charge state. Further, these measurements also exclude all potential background caused by the setup components (type~B), which would be constant throughout the measurements.\\
Thorium atomic shell effects such as a long-lived atomic excitation or a chemical reaction between thorium and the MCP surface could potentially contribute background  (type~C). To exclude this possibility it is sufficient to perform a comparative measurement with $^{230}$Th where such effects would be identical (Extended Data Table~1 no.~3). For this purpose, a $^{234}$U source was employed (270~kBq, electrodeposited onto a titanium sputtered silicon wafer, in the following denoted as source 2). The $^{230}$Th $\alpha$-recoil ions emerging from this source were accumulated on the surface of the MCP detector for 2,000 s, just as for $^{229}$Th. For $^{230}$Th, however, no signal is detected (Fig.~4~b, second row), which proves that the signal obtained for $^{229}$Th cannot be caused by an atomic shell effect. This measurement also provides further exclusion of
\end{multicols*}
\begin{figure}[H]
\begin{center}
\vspace{-2.5cm}
 \includegraphics[width=18.3cm]{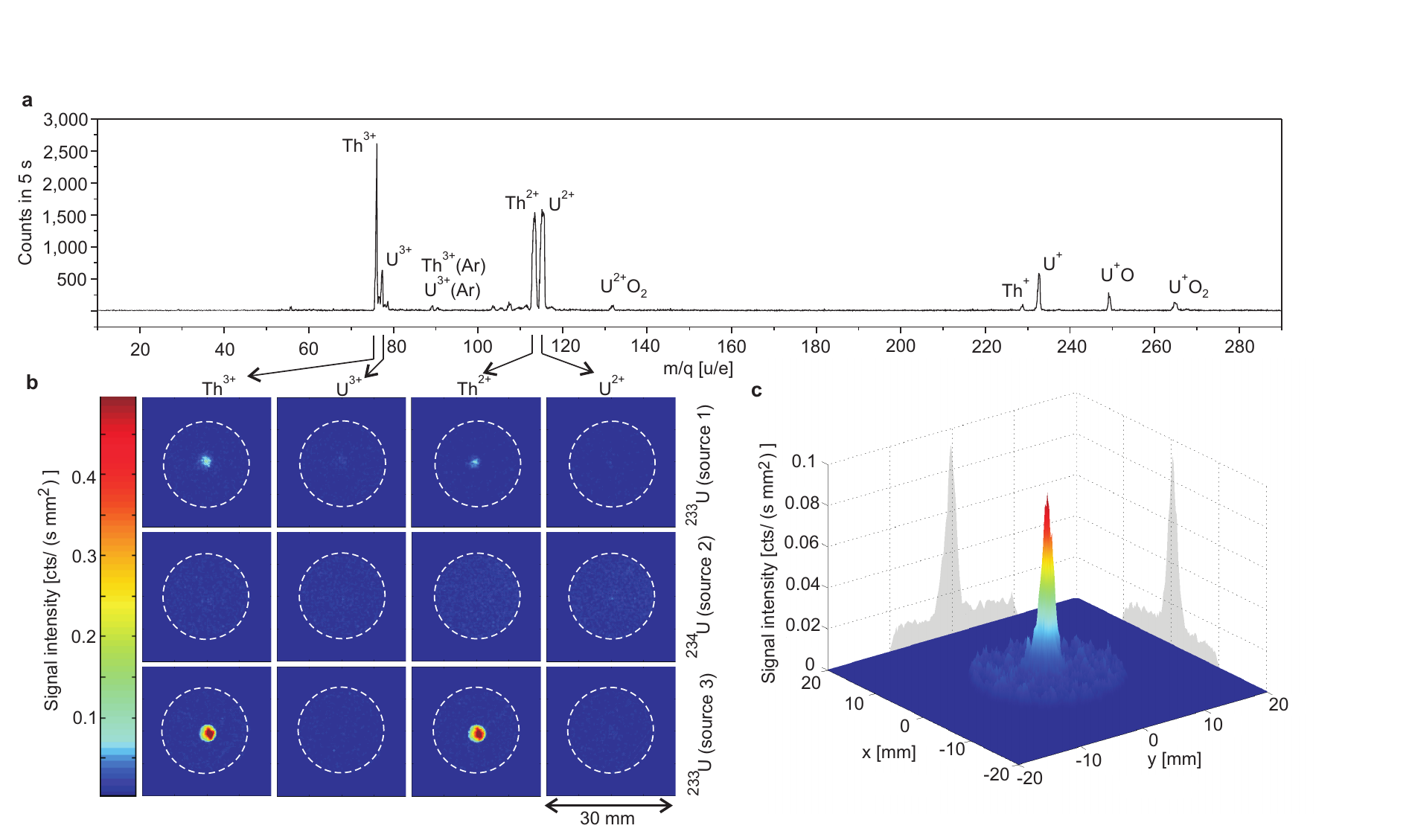}
  \caption{\small {\bf Signal comparison.} a: Complete mass scan performed with the $^{233}$U source~1 \cite{Wense2}. b: Comparison of MCP signals as obtained during accumulation of thorium and uranium in the $2+$ and $3+$ charge states, originating from $^{233}$U and $^{234}$U sources, respectively. Each image corresponds to an individual measurement of 2,000~s integration time (20~mm diameter aperture indicated by dashed circles). The rows correspond to different uranium sources, while each column corresponds to an individual extracted ion species as indicated by the arrows from the mass scan. Measurements were performed at about $-25$~V MCP surface voltage in order to guarantee soft landing of the ions. c: Signal of the $^{229}$Th isomeric decay as obtained during $^{229}$Th$^{3+}$ extraction with source~1. A signal area diameter of about 2~mm (FWHM) is achieved. The obtained maximum signal intensity is 0.08~cts/s mm$^2$ at a background rate of about 0.01~cts/s~mm$^2$.} 
 \label{allimages}
 \end{center}
\end{figure}
\begin{multicols*}{2}
\noindent background of types (A) and (B). In this way most of the systematic background effects are excluded.\\ 
In former experiments, direct identification of the $^{229\mathrm{m}}$Th isomeric decay has been prevented in part by radioactive decay of short-lived daughter nuclides \cite{Peik3}. Our experiments focused specially on this type of potential background (type~D), that we have been able to exclude in four independent ways. A quadrupole mass-separator (QMS) is used for the extraction of ions with a well-defined mass-to-charge ratio from the buffer-gas stopping cell. The achieved mass-resolving power of $m/\Delta m=150$ is sufficient for the complete separation of the $\alpha$-recoil ions with a difference of four or more atomic mass units (Extended Data Fig.~3). Figure~5~b shows the signal intensity as a function of the selected mass-to-charge ratio $m/q$ for MCP surface voltages of $-25$~V and $-2,000$~V. At a $-2,000$~V surface voltage (blue) the ion-impact signal is observable and the $^{233}$U$^{2+}$ and $^{229}$Th$^{2+}$ mass peaks are of comparable amplitude. At the $-25$~V surface voltage (red) the $^{233}$U$^{2+}$ mass peak completely vanishes, since no ion-impact signal is detected. $^{229}$Th$^{2+}$, in contrast, reveals a remaining component, which is clearly restricted to the $^{229}$Th$^{2+}$ mass peak. However, molecular sidebands may be populated by nuclides of lower masses (e.g. $^{213}$Bi$^{16}$O reveals the same mass as $^{229}$Th and is a $\beta^-$ emitter in the $^{233}$U decay chain with a 45.6 minute half-life). Thus restriction to the $m/q$ value of $^{229}$Th$^{2+}$ does not exclude short-lived daughter nuclides as signal contributions.\\
One way to exclude this sort of background is obtained from the parallel observation of the signals in the $2+$ and the $3+$ charge states (Extended Data Table~1 no.~4, Fig.~4~b, first row), because only thorium is extracted to a significant amount in the $3+$ charge state due to its low 3$^{\mathrm{rd}}$ ionisation potential \cite{Wense2} (see Extended Data Table~2 for comparison). Experimentally, a suppression of three to four orders of magnitude for the short-lived daughters in the $3+$ charge state compared to the $2+$ charge state was obtained \cite{Wense2}.\\
One further comparison (Extended Data Table~1 no.~5) was performed with a newly available chemically purified $^{233}$U source (source~3, 290~kBq, same geometry as source~2). The factor of chemical purification of the short-lived daughter nuclides was measured to be $\ge250$. In case signals were originating from nuclear background, a drastically reduced signal intensity should occur. This reduction is, however, not observed and instead the signal increases by a factor of $\sim13.5$ due to a larger $^{233}$U content and a reduced source thickness, leading to a higher $\alpha$-recoil efficiency (Fig.~4~b, third row).\\
Two other ways for excluding nuclear background are discussed in the Methods section. Consequently, the nuclear isomeric transition in $^{229}$Th is the only possible explanation for the observed signal.\\[0.5cm]
\begin{large}
\noindent{\bf \textcolor{blue}{Half-life and energy constraints}}\\[0.2cm]
\end{large}
Direct detection of the $^{229\mathrm{m}}$Th isomeric-decay signal provides first constraints on the half-life of the isomer, which is found to be heavily charge-state dependent. Two different
\end{multicols*}
\begin{figure}[H]
 \begin{center}
 \vspace{-1.5cm}
 		\includegraphics[width=18.3cm]{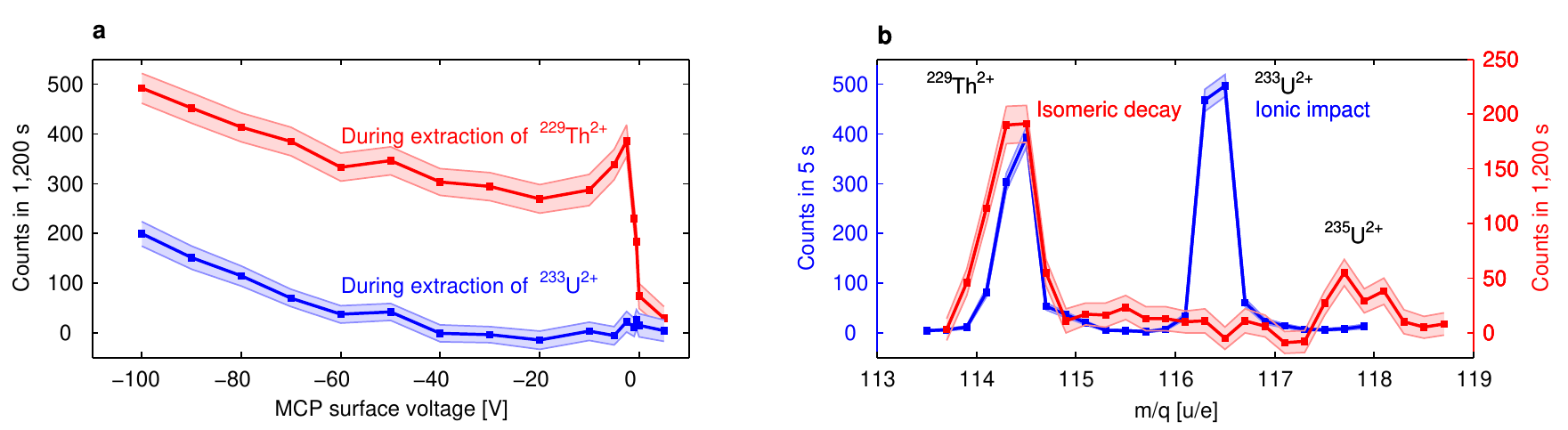}
 \vspace{0cm}
  \caption{\small {\bf Background corrected $^{229}$Th isomeric decay signals.} a: $^{229}$Th$^{2+}$ signal (red) in comparison to $^{233}$U$^{2+}$ (blue) as a function of the MCP surface voltage. The errors are indicated by shaded bands. They are given as the estimated standard deviation of the poisson distribution for sample size $n=1$ ($\sqrt{S+N}$, where $S$ and $N$ denote the total and the background count numbers of about 60 counts, respectively). 
b: Signal of extracted ions as a function of the mass-to-charge ratio behind the QMS for MCP surface voltages of $-25$~V (isomeric decay, red) and $-2,000$~V (ion impact, blue). Note the different integration times and axes scales. Besides the signal at 114.5~u/e (corresponding to $^{229}$Th$^{2+}$), a further signal at 117.5~u/e occurs, which originates from the isomeric decay of $^{235}$U ($^{239}$Pu was shown to be contained in the source material by $\alpha$ spectroscopy \cite{Wense2}, the isomer is populated by a 70\% decay branch and the extraction rate is too small to be visible in the ion-impact signal).}
 \label{massdep}
 \end{center}
\end{figure}
\vspace{-1cm}
\begin{multicols*}{2}
\noindent measurements were performed (see Methods for details): the first one conducted to estimate an upper limit for the isomeric half-life in the neutral thorium atom, and the second one conducted to infer a lower limit for the isomer's lifetime in $^{229}$Th$^{2+}$. These measurements allow to draw conclusions on the isomeric energy, as the half-life changes depending on whether the IC decay channel is energetically permitted or not.\\
For the neutral thorium atom, $^{229\mathrm{m}}$Th is predicted to decay predominantly by IC with a half-life as short as microseconds \cite{Karpeshin1,Tkalya4}.  Experimentally an upper limit for the isomeric half-life in neutral thorium was found by $^{229\mathrm{(m)}}$Th$^{2+}$ ion-beam pulsing. Images were acquired directly after the ion-pulse had impacted the MCP surface, leading to the formation of neutral thorium by charge exchange. In this way the half-life was determined to be less than one second, confirming that the isomeric IC decay-channel is energetically allowed. This in turn gives strong indication that the isomeric energy is above the 1$^{\mathrm{st}}$ ionisation potential of thorium of 6.31~eV.\\
An isomeric half-life of minutes to hours has been predicted for $^{229\mathrm{m}}$Th in a charge state $\ge 1+$, where IC is energetically forbidden \cite{Karpeshin1,Tkalya4}. In order to confirm this prediction, $^{229\mathrm{(m)}}$Th$^{2+}$ ions were stored in the RFQ before acquiring the isomeric-decay signal. The half-life range probed in this way was limited by the maximum ion storage time in the RFQ, which is about 60~s. Still, after this time, significant isomeric decay was detected, suggesting the isomeric lifetime in Th$^{2+}$ to be longer than 60~s. This long half-life can only be explained if the isomeric IC decay-channel is energetically forbidden for $^{229}$Th$^{2+}$. Thus the isomeric energy must be below the 3$^{\mathrm{rd}}$ ionisation potential of thorium of 18.3~eV.\\
Based on the half-life estimates, the value of the isomeric energy is deduced to be between 6.3 and 18.3~eV (i.e. between the 1$^{\mathrm{st}}$ and 3$^{\mathrm{rd}}$ ionisation potential of thorium). This energy range is consistent with today's most accepted value \cite{Beck1} and promising for the development of a nuclear clock based on thorium ions.\\[0.5cm]
\begin{large}
\noindent{\bf \textcolor{blue}{Discussion and perspectives}}\\[0.2cm]
\end{large}
The efficient production of a low-energy, highly pure $^{229\mathrm{(m)}}$Th ion beam enabled the successful direct observation of the $^{229}$Th isomer to ground-state decay, via spatially decoupled isomer population and isomeric decay, combined with an efficient mass separation using a QMS.\\
This measurement is not only a further proof of the isomer's existence, which has been controversial \cite{Matinyan,Sakharov}, but also provides a detection method that can be used as a tool to probe different processes for isomer population, e.g., via direct laser excitation \cite{Campbell1} or electronic bridge processes \cite{Peik3}. Further, in the nuclear-clock concept the observed IC decay could be used to probe the isomeric population to provide an alternative to the double-resonance method proposed so far \cite{Peik1}. Most importantly, this direct detection paves the way for precise determination of the isomer's decay parameters. The isomeric half-life can be probed by applying a cryogenically cooled Paul trap \cite{Schwarz}, which allows for longer ionic storage times. A significantly more precise energy value can be determined by applying a hemispherical electron energy analyser \cite{Martensson} with an energy resolution of a few meV (see Methods for details). This will usher the possibility for developing a laser system that can ultimately bring all-optical control of this nuclear transition and thus provide a template for coherent manipulation of nuclei in general \cite{Liao}. The construction of a nuclear frequency standard based on this $^{229}$Th isomeric transition will enable new perspectives for ultra-precise frequency metrology and is expected to have implications for both technology and fundamental physics.\\[0.5cm]

\begin{small}
\bibliographystyle{nature}

\noindent{\bf Acknowledgements} We acknowledge fruitful discussions with D. Habs, T.W. H\"ansch, T. Udem, T. Lamour, J. Weitenberg, A. Ozawa, E. Peters, J. Schreiber, P. Hilz, T. Schumm, S. Stellmer, F. Allegretti, P. Feulner, J. Crespo, M. Schwarz, L. Schm\"oger, P. Micke, C. Weber, P. Bolton and K. Parodi. We thank T. Lauer for the Ti-sputtering of the Si-wafers and the MPQ for the temporary loan of the MCP detector. We thank I. Cortrie L. Black and J. Soll for graphic design support. This work was supported by DFG (Th956/3-1) and by the European Union's Horizon 2020 research and innovation programme under grant agreement No 664732 "nuClock". \\
\noindent{\bf Author Contributions} L.v.d.W., B.S. and P.G.T. performed the experiments. M.L. and J.B.N. did preparatory experimental work. H.-J.M. and H.-F.W. produced the radioactive source 1. C.M., J.R., K.E., C.E.D., N.G.T. and L.v.d.W. produced the radioactive sources 2 and 3. L.v.d.W., P.G.T. and B.S. wrote the manuscript with input from all authors.\\
\noindent{\bf Author Information} Reprints and permissions information is available at www.nature.com/reprints. The authors declare no competing financial interests. Readers are welcome to comment on the online version of the paper. Correspondence and requests for materials should be addressed to L.v.d.W.(L.Wense@physik.uni-muenchen.de).
\end{small}

\newpage
\begin{large}
\noindent{\bf \textcolor{blue}{Methods}}\\[0.2cm]
\end{large}
\begin{small}
\textbf{$^{233}$U and $^{234}$U $\alpha$-recoil ion sources.} Three different sources were employed in these experiments. Source 1 consists of about 200~kBq $^{233}$U (UF$_4$), evaporated in vacuum from a tantalum heater lined with a vitreous carbon crucible \cite{Maier} onto a 20-mm diameter stainless-steel plate. The preparation was performed in the former hot-lab facility of the LMU Munich \cite{Grossmann}. The UF$_4$-layer thickness is (360$\pm$20)~nm, leading to a recoil efficiency of about 5.3\% for $^{229}$Th. The source material was not chemically purified before evaporation. As the material was produced around 1969, a significant ingrowth of short-lived daughter nuclides occurred since then. An unavoidable fraction of $^{232}$U contamination was determined by $\gamma$ spectroscopy to $(6.1\pm0.3)\cdot 10^{-7}$ at the time of material production \cite{Wense2}.\\
Source 2 consists of ($270\pm10$) kBq $^{234}$U, deposited by molecular plating \cite{Vascon} onto the surface of a Ti-sputtered Si-wafer of 100~mm diameter. It has a thickness of 0.5~mm with a 100~nm thick layer of sputtered titanium. The active surface area of the source is 90 mm in diameter, leaving a 12~mm diameter unplated region in the center.\\
Source 3 is a newly available $^{233}$U source of about 290~kBq. Just like source 2, it was deposited by molecular plating with 90 mm diameter onto the surface of a Ti-sputtered Si-wafer of 100 mm diameter. Due to the smaller source thickness, the thorium extraction rate was improved by a factor of about 13.5 compared to source 1. The source 3 material was chemically purified before deposition by ion-exchange chromatography to remove the $^{233}$U and $^{232}$U daughter nuclides. A relative purification factor of $\ge250$ was found, based on a comparison of $\gamma$-energy spectra of the source material before and after chemical purification.\\
\textbf{Buffer-gas stopping cell.} The uranium source is mounted into the buffer-gas stopping cell \cite{Neumayr1} (Extended Data Fig.~2) and acts as an electrode of the ion-extraction system (39~V offset voltage). The $\alpha$-recoil ions, which possess a kinetic energy of up to 84.3~keV for $^{229}$Th, are stopped in 40~mbar of ultra-pure helium. In order to guarantee the required cleanliness of the buffer gas, helium with a purity of 99.9999\% is used, which is further purified by catalytic purification (SAES Getters, MonoTorr, phase 2) and a cryotrap filled with liquid nitrogen. The gas tubing was electropolished and the cell chamber was built to UHV standards, bakeable up to 180°C. A typical background pressure of $p \le 3\cdot 10^{-10}$~mbar is achieved. This high cleanliness allows for the extraction of $^{229}$Th even in the 3+ charge state \cite{Wense2}.\\
The buffer-gas stopping cell also houses the RF+DC-funnel system, consisting of 50 ring electrodes of 0.5 to 1~mm thickness, converging from 115~mm to 5~mm inner diameter. RF- and DC voltages are applied to this electrode structure. The applied RF voltages are 220~V$_{\mathrm{pp}}$ at 850~kHz, varying in phase by 180° between neighbouring electrodes. This leads to a repelling force, preventing the recoil ions from charge exchange at the electrodes. In parallel, a DC voltage gradient of 4~V/cm is applied by a voltage-divider chain (35 to 3~V), guiding the ions through the buffer-gas background towards the buffer-gas stopping-cell exit.\\
The latter consists of a supersonic Laval-nozzle (2~V offset) with a 0.6~mm diameter nozzle throat. In this way, supersonic velocities of the helium gas flow are achieved and the $\alpha$-recoil ions are extracted from the buffer-gas stopping cell together with the helium carrier gas.\\
\textbf{RFQ ion guide and cooler.} Following the buffer-gas stopping cell, the ions are injected into an RFQ system, which consists of four rods with 11~mm diameter, with a 10~mm distance between opposite rods. For ion guiding an RF field of 200~V$_{\mathrm{pp}}$ at 880~kHz is applied. Each rod is divided into 12 segments and the overall length of the system is 33~cm. Because of the segmentation we can apply an individual DC voltage to each segment, thereby establishing a voltage gradient of 0.1~V/cm (1.8 to 0~V) to drag ions through the remaining helium buffer-gas background of about $10^{-2}$~mbar, or to store the ions in the RFQ. This background pressure is used for phase-space cooling of the recoil ions, which leads to a sub-mm diameter recoil-ion beam at the RFQ exit. By voltage control of the last RFQ electrode the ion beam can optionally be pulsed.\\ 
\textbf{Quadrupole mass-separator.}  Following the RFQ, the $\alpha$-recoil ions are mass separated in a quadrupole mass-separator (QMS) \cite{Haettner}. The QMS consists of four rods with 18 mm rod diameter and 15.96~mm inner rod distance. The length is 30~cm, with an additional 5~cm at the entrance and exit acting as Brubaker lenses \cite{Brubaker}. At the resonance frequency of 925~kHz an RF amplitude of 600.5~V$_\mathrm{pp}$ and a DC voltage of 50.15~V is required for the extraction of $^{229}$Th$^{3+}$ (901.5~V$_\mathrm{pp}$ and 75.23~V for the $2+$ charge state, respectively). A voltage offset of $-2$~V is applied to the whole system. With this device, a transmission efficiency exceeding 70\% with a mass resolving power of $m/\Delta m = 150$ can be achieved.\\
Prior to any isomer detection, the QMS is calibrated in order to extract ions of wanted mass-to-charge ratio. The mass spectrum (Fig.~4~a) is well known from earlier measurements \cite{Wense2}, where the correctness of the peak assignment was proven by parallel detection with a silicon detector for $\alpha$ spectroscopy and an MCP detector. Given this mass spectrum, the QMS is calibrated by performing ion-impact profile measurements (Extended Data Fig.~3 lower panel) with the beam-imaging MCP detector (Beam Imaging Solutions, BOS-75-FO), when operating the detector at a surface voltage of about $-900$~V. Consequently the impact of the transmitted ions is detectable (due to their kinetic energy of 1.8 to 2.7~keV, depending on the charge state). During calibration, care has to be taken not to contaminate the detector surface with short-lived daughter nuclides. For this purpose, the scans are always started at higher masses (above $^{233}$U) and stopped when the $^{229}$Th$^{2+}$ mass peak is reached.\\
\textbf{Triodic extraction system.} Behind the QMS, the ions are guided by a triodic electrode structure consisting of three ring electrodes in a nozzle-like shape: The first electrode acts as an aperture electrode to shield the RF voltages of the QMS ($-2$~V). A voltage of $-62$~V is applied to the second electrode in order to extract the ions from the QMS. The third electrode with a 2-mm diameter opening shields the extraction voltage from the surrounding when applying $-22$~V. As a result ions are guided to the MCP detection system. A combined extraction and purification efficiency for Th$^{3+}$ of $(10\pm2)$\% was determined behind the triodic extraction system \cite{Wense2}. Together with the 5.3\% recoil efficiency of source~1, $(1.0\pm0.1)\cdot 10^3$ $^{229}$Th$^{3+}$ ions per second are extracted. A $(5.5\pm1.1)$\% extraction efficiency was obtained for Th$^{2+}$, resulting in $(5.8\pm0.6)\cdot 10^2$ extracted Th$^{2+}$ ions per second. The total time for extraction is a few ms (3 to 5~ms were obtained as extraction time behind the RFQ \cite{Neumayr2}). Faster decays of nuclear excitations already take place in the buffer-gas stopping cell.\\ 
\textbf{MCP detection system.} The ions are collected directly on the surface of a micro-channel plate (MCP) detector \cite{Wiza} placed at 5~mm distance to the last electrode of the triodic extraction system (Fig.~2). The MCP detector (Beam Imaging Solutions, BOS-75-FO) consists of two MCP plates (chevron geometry, 25~$\mu$m channel diameter) with 75~mm diameter. The front surface is CsI-coated. The two plates are positioned in front of a vacuum-flange-mounted optic fibre-glass window, which is coated with a phosphor layer. During extraction, the MCP is operated at a He pressure of $10^{-6}$~mbar and typical voltages of $-25$~V and $+1,900$~V are applied to the front and the back sides of the MCP, respectively. A voltage of $+6,000$~V is applied to the phosphor screen, which is monitored through the optic fibre-glass window by a CCD camera (FL2-14S3M-C, PointGrey) with a zoom lens (Computar M2514MP2, 25~mm, C-mount). The distance between the window and the CCD camera is about 30~cm, leading to a field of view of 100~mm by 75~mm. The outer region of the optical window is covered by a 20-mm diameter aperture in order to cover arcing effects from the detector's side. The camera is mounted onto an optical rail, which is placed in a light-tight housing.\\
Due to the expected short isomeric lifetime in neutral thorium, it is important to allow for $^{229\mathrm{m}}$Th decay detection during ion accumulation, which affords probing even for decays that would occur simultaneously with charge exchange on the MCP surface. For this purpose, the MCP is operated with a surface voltage of $-25$~V. In this way the thorium ions are collected at low kinetic energy (50-75~eV, depending on the charge state) in soft landing onto the MCP surface. The remaining kinetic energy of the ions as well as the energy carried by the ions in form of the charge state does not lead to a significant signal on the MCP surface \cite{Rispoli}. Most of the energy in these processes is transferred to phonons at the point of impact with the surface \cite{Bay}. No ion-impact signal was detected with an MCP surface voltage above $-40$~V (i.e. negative voltage with magnitude below 40~V).\\  
Relatively little is known about the detection efficiency of MCPs for low-energy electrons (the ionisation potential of thorium is 6.31~eV, thus an IC electron kinetic energy of about 1.5~eV remains, given a 7.8-eV isomeric transition). Applying the model discussed in \cite{Goruganthu}, a decrease in detection efficiency to 2.9\% of the maximum value (at about 300~eV kinetic energy) is predicted for incident electrons of 1.5~eV energy. Assuming a maximum detection efficiency of 50\% (corresponding to channel open area of the MCP), an absolute detection efficiency of about 1.5\% is expected. 2\% of the 1,000 $^{229}$Th$^{3+}$ ions which are extracted per second are predicted to be in the isomeric state \cite{Barci}. Comparing this with the detected isomeric-decay count rate of 0.25 per second leads to an experimentally obtained detection efficiency of 1.3\%, which is in good agreement with our expectation.\\
\textbf{Image readout.} For readout of the MCP signal, the CCD-chip (Sony ICX267 CCD, 4.65$\times$4.65~$\mu$m$^{2}$ pixel size, 1384$\times$1032 pixels) was exposed for 4~s for each frame. In these frames, single events of the MCP detector can clearly be distinguished from the CCD intrinsic background (noise and hot pixels) by size and intensity. A Matlab program is applied to determine the position of each individual event. These events are then added for a chosen number of frames (typically 500 for 2,000 s integration time) to obtain one single image. Appropriate choice of the filter parameters of the program is tested by an individual control of 50 images. The loss of events due to low signal intensity on the phosphor screen or due to spatial overlap is found to be negligible. Only a minor amount of CCD intrinsic noise is not adequately filtered. By applying this type of image readout, the background is dominated by the MCP intrinsic dark-count rate of about 0.01~cts/s~mm$^2$.\\
\textbf{Signal comparison.} CCD camera images of the phosphor screen reveal features that enable to distinguish type or origin of signals. Signals of different origin are shown in Extended Data Fig.~4. Each image corresponds to 4~s exposure time of the CCD chip (i.e. one frame). A wanted ion species was chosen by mass-to-charge selection with the QMS. Extended data Fig.~4~a shows $\alpha$ decays on the MCP surface occurring within 5~min after extraction of $^{221}$Fr$^{2+}$ ($t_{1/2}=286$~s). Very large and intense signals are seen, with an average diameter of about 1~mm. Extended Data Fig.~4~b shows $\beta$ decays occurring within 45~min after extraction of $^{209}$Pb$^{2+}$ ($t_{1/2}=3.25$~h). The signals are significantly smaller and less intense than the ones caused by $\alpha$ decays. The typical signal diameter is about 0.6~mm. Extended Data Fig.~4~c shows signals caused by the isomeric decay of $^{229}$Th starting to occur instantaneously with the accumulation of $^{229}$Th$^{3+}$ on the MCP surface. The signals appear small and of low intensity with a typical signal diameter of about 0.3~mm. They are slightly smaller compared to the signals caused by $\beta$ decays, and clearly distinguishable from the $\alpha$ events. Finally, signals caused by the isomeric 76~eV IC decay of $^{235\mathrm{(m)}}$U ($t_{1/2}=26$~min) are shown in Extended Data Fig.~4~d, taken within 30~min of extraction of $^{235}$U$^{2+}$. They are comparable with the isomeric-decay signals of $^{229}$Th.\\
\textbf{Half-life measurements.} Two different half-life measurements are implemented. The first measurement leads to an upper limit for the isomeric half-life in neutral thorium. To obtain this limiting value, a pulsed $^{229\mathrm{(m)}}$Th$^{2+}$ ion beam is produced by applying a gate voltage of 0.5~V to the last RFQ electrode. The gate is opened for 500~ms and is then closed for 1700~ms, while ions are accumulated in the RFQ continuously (a maximum storage time of about 1 minute is obtained for Th$^{2+}$). Strong ion pulses are produced when the QMS is set to extract $^{229}$Th$^{2+}$. This is controlled by applying an MCP surface voltage of about $-900$~V, yielding strong ion-impact signals. The CCD camera acquires images of 1~s exposure time only when the beam gate is closed. To ensure that the gate is actually closed, the camera is started 500~ms after applying the gate voltage. The camera is stopped after 1,200~ms in parallel to the gate opening, in order to acquire one image per pulse. It is reconfirmed that the camera does not acquire pictures at times of ion impact. By the sequence 1,200 frames (corresponding to 1,200~s total exposure time) are evaluated. No signal is obtained, which means that the isomer half-life must be below 1 second, allowing for charge exchange of the $^{229}$Th$^{2+}$ ions on the MCP surface. \\
In a second measurement, a lower limit of the isomeric lifetime in $^{229}$Th$^{2+}$ is found. For this purpose, $^{229\mathrm{(m)}}$Th$^{2+}$ ions are stored in the RFQ, by applying a gating voltage of 5~V to the last RFQ electrode. After storage, the ion cloud is accelerated onto the MCP surface to examine survival of the isomeric state by detected internal conversion. The half-lives, that can be probed by this method are limited by the storage times of Th$^{2+}$ in the RFQ. A one minute storage time is easily accessible without significant ion loss. For this measurement the ions are accumulated for 10~s in the RFQ, where they are stored. After 10 seconds the $^{233}$U source offset is reduced to 0~V, preventing additional recoil ions from leaving the buffer-gas stopping cell. Then the ions are stored for one minute in the RFQ, waiting for the isomeric decay to occur. Afterwards, the gate voltage is also reduced to 0~V and the isomeric decay is read from the MCP detector. To reduce the dark count, the CCD camera is triggered to only acquire images when the ions are released. In this way, 200 pulses are evaluated with 3 imaged frames per pulse (4~s exposure time for each frame). A clear signal is seen when the QMS is set to extract $^{229}$Th$^{2+}$, from which is inferred a half-life greater one minute. To eliminate signal contribution from a long-lived $\beta^-$ emitter, which might have populated the $^{229}$Th$^{2+}$ mass peak by molecular formation (e.g. oxides), a measurement of the background rate is performed afterwards for 1 hour and no signal is obtained.\\
\textbf{Exclusion of nuclear background based on signal comparison between $^{229}$Th$^{2+}$ and $^{229}$Th$^{3+}$} (Extended Data Table.~1 no.~4)\textbf{.}  All potential background contributions together with the relevant means of exclusion are listed in Extended Data Table.~1. The given measurement numbers refer to this figure.\\
It was discussed that the parallel occurrence of the signal in the $2+$ and $3+$ charge states (Fig.~4~b, first row) is already sufficient to exclude nuclear background as potentially originating from short-lived daughter nuclides. The reason is that only thorium can be expected to be extracted to a significant amount in the $3+$ charge state, due to its low 3$^{\mathrm{rd}}$ ionisation potential of only 18.3~eV (see Extended Data Table~2), which is below the first ionisation potential of He (24.6~eV). Thus, during stopping in the helium environment, it is energetically favourable for the electrons to stay attached to the helium atoms instead of reducing the thorium $3+$ charge state. Experimentally, a reduced extraction for all short-lived daughter nuclides (of atomic number $Z=88$ or below) by three to four orders of magnitude is found in the $3+$ compared to the $2+$ charge state \cite{Wense2}. In case of signals not caused by $^{229}$Th, the same reduction of signal intensity would be expected when comparing the $2+$ and $3+$ charge states. This, however, is not observed. For completeness, all ionisation potentials \cite{ionize} for the elements which are potentially contained in the source material are listed in Extended Data Table~2. Also the heavier elements possess low ionisation potentials, however, they cannot explain the observed signal, because their half-lives are significantly longer. Further, mass-peak shifts from heavier to lighter masses cannot be explained by the population of molecular side-bands.\\
\textbf{Exclusion of nuclear background based on signal comparison between $^{229}$Th originating from chemically purified and unpurified $^{233}$U sources} (Extended Data Table~1 no.~5)\textbf{.}
Isomeric-decay measurements were also performed with the new chemically purified $^{233}$U source of 90~mm diameter (source~3, 290~kBq). The thorium ions were collected in the $2+$ and $3+$ charge states for 2,000~s, while detection was performed in parallel. Compared to source~1, these measurements resulted in $\sim13.5$ times higher isomeric count rate ($\sim 3.4$~cts/s). This enhancement occurs due to the reduced source thickness, leading to a higher $\alpha$-recoil efficiency. The results of these measurements are shown in Fig.~4~b, third row.\\
If signals were caused by a decay of any of the short-lived daughter nuclides, a significant decrease in signal intensity compared to source 1 should occur, due to the chemical purification factor of more than 250. The fact that this is not the case further serves to exclude radioactive decay of short-lived isotopes as a signal contribution.\\
\textbf{Exclusion of nuclear background based on signal appearance and the $^{229\mathrm{m}}$Th half-life limit} (Extended Data Table~1 no.~6)\textbf{.}  It was shown that the uniquely strong signal shape excludes any $\alpha$ decay as contributor to the observed decay events (see Extended Data Fig.~4). This information, in combination with the observed short decay half-life (in the sub-second region), is already sufficient to exclude any nuclear origin of the signal except for the isomeric decay of $^{229}$Th.\\
 While the uranium-source material predominantly consists of $^{233}$U, in source 1, trace amounts of other nuclides are also included ($^{232}$U, $^{238}$Pu, $^{239}$Pu, $^{231}$Pa) together with their decay daughters \cite{Wense2}. Even further nuclides could potentially be present, although they have not been experimentally observed. A complete list of nuclides potentially contained in the source material (produced by neutron irradiation in a nuclear reactor) is shown in Extended Data Fig.~5. Also their half-lives and decay branching ratios are listed \cite{gov}. For completeness, all populated nuclides are shown, even if their activity can be assumed to play only a negligible role due to their small branching ratio or due to a long half-life of the mother nuclide. A complete list of potentially contributing isomers is given in Extended Data Table~3, together with their corresponding excitation energies and half-lives \cite{gov}. Note that excited states with half-lives in the $\mu$s range or shorter do not have to be considered, because the extraction time from the source is in the ms range \cite{Neumayr2}.\\
As can be inferred, there is no pure $\beta$ emitter or isomer contained, which could potentially explain the detected signal, except for the 0.8~s isomeric state in $^{207}$Pb. This isomeric state, however, is populated only by a fraction of $8.1 \cdot 10^{-6}$ from the $\alpha$ decay of $^{211}$Po, which itself is not part of the main decay chains.\\
Furthermore, the $^{229}$Th isomeric transition is the only known nuclear transition which is expected to reveal the observed strong dependence of its half-life on the electronic environment. Thus the detected signal cannot be explained by any nuclear decay other than the decay of the $^{229}$Th first excited nuclear state.\\
\textbf{Exclusion of nuclear background by search for $\alpha$ and $\beta$ decays using Si and LN$_2$ cooled Si(Li) detectors} (Extended Data Table~1 no.~7)\textbf{.} To further substantiate the evidence, the extracted ions (when operating the QMS for collection of $^{229}$Th$^{2+}$ or $^{229}$Th$^{3+}$) are directly accumulated on the surface of two different silicon detectors. The first detector is optimised for $\alpha$-particle detection in order to provide a further exclusion of $\alpha$ decays as a cause of the detected signals. The second detector is used in order to exclude $\beta$ decays or high-energy internal conversion electrons.\\
For the exclusion of $\alpha$ decays an ion-implanted silicon charged particle detector (Ametek, BU-014-150-100) is used. This detector is mounted directly behind the extraction triode with about 5~mm distance, replacing the CsI-coated MCP detector. A charge sensitive preamplifier (CSTA) and a shaping amplifier (Ortec, model 571) are used for signal processing. The spectra are acquired by a multi-channel analyzer (Amptek, MCA-8000A). The detector is operated at a 20~V bias voltage and a $-10$~V offset is applied to the whole system in order to collect the ions directly on the detector surface. To allow also for mass scans as required for the calibration of the QMS an MCP detector (Hamamatsu, type F2223) is mounted sideways at 90° to the extraction triode. During QMS calibration, a surface voltage of $-2,000$~V is applied to the MCP, which is sufficiently high to attract the ions in spite of its off-axis position. After the QMS has been set to extract the desired ion species, the MCP surface voltage is reduced to $0$~V, so ions are collected on the Si-detector surface. In this way, four different measurements were performed each with 2 hours acquisition time: one during the extraction of $^{213}$Bi$^{2+}$ (2.0~cts/s) in order to prove the functionality of the detector system, one dark count measurement ($5.7\cdot 10^{-3}$~cts/s), one during the extraction of $^{229}$Th$^{2+}$ ($6.0\cdot 10^{-3}$~cts/s) and one during the extraction of $^{229}$Th$^{3+}$ ($5.3\cdot 10^{-3}$~cts/s). The corresponding spectra are shown in Extended Data Fig.~6 a-d. As expected, no entries above background are visible in the energy range where $\alpha$-particles from $^{229}$Th would appear (4.7-5.1 MeV) during the extraction of $^{229}$Th in the $2+$ or $3+$ charge state, as the half-life of the $^{229}$Th $\alpha$ decay is 7932 years and thus practically no decays occur within the duration of these comparatively short measurements. The fact that no line from any $\alpha$-decaying nucleus is visible in the spectra allows excluding $\alpha$-decays as the origin of the ~0.25 cts/s-signal measured on the MCP in the search for the isomeric decay of $^{229\mathrm{m}}$Th: had this signal originated from $\alpha$-decays, a total of about 1,800 cts should be seen in a 2 hour measurement with the Si detector, which would have been easily visible.\\
While $\beta$ decays as a signal contribution have already been excluded by half-life arguments, a further way to exclude them is given by direct $\beta$ detection. For this purpose a liquid nitrogen cryogenically cooled Si(Li) detector (Canberra, type ESLB-3000-300) is used, replacing the above mentioned Si detector. It is operated in combination with a preamplifier with a cooled FET stage (Eurisys Measures, PSC 761) and a shaping amplifier (Ortec, model 572). Again spectra are acquired by a multi-channel analyzer (Amptek, MCA-8000A). A bias voltage of $-400$ V is applied to the front surface, such that no further offset is required. The detector is mounted at 5~mm distance to the triodic extraction system. Four different measurements were performed, each with 10 hours acquisition time: one dark-count measurement (0.47~cts/s), one during the extraction of $^{229}$Th$^{2+}$ (0.44~cts/s), one during the extraction of $^{229}$Th$^{3+}$ (0.48~cts/s) and one during the extraction of $^{209}$Pb$^{2+}$ (2.13~cts/s), the latter to prove the functionality of the detection system. If the detected signals were $\beta$ decays or high-energy internal-conversion electrons, the expected enhancement of the integrated signal rate of (0.25$\pm$0.1) cts/s (for source 1) would have been detected easily.\\
\textbf{Prospects for energy determination.} A precise determination of the isomer's energy is one of the most important prerequisites for the development of a nuclear frequency standard. The direct detection of the isomeric decay opens new perspectives for such an energy determination. In the presented work, the IC decay channel in the neutral thorium atom is investigated. Any energy determination based on this direct detection will require energy spectroscopy of the IC electrons emitted in the isomeric decay.\\
Several techniques for electron-energy spectroscopy of different precisions and complexities are known. The highest known precision is provided by hemispherical electron energy analysers, which possess resolutions in the range of a few meV \cite{Martensson}. While being also among the most complex devices for spectroscopy, there is a tradeoff between energy resolution and signal contrast. This problem can be solved by ion-beam pulsing. When applying an RFQ buncher, ion bunches of a few 10~ns pulse length can be produced \cite{Plass}. These bunches are significantly shorter than the expected isomeric lifetime in the neutral thorium atom, which is predicted to be in the $\mu$s range. Such ion beam pulsing would not only allow to determine the isomer's half-life in the neutral thorium atom, but also to suppress any background by several orders of magnitude (depending on the exact isomeric half-life) if the electron detector is triggered in accordance with these pulses. This improvement in signal-to-background ratio will make high resolution electron spectroscopy applicable to the problem of energy determination of the isomeric state.\\
A significantly simpler sort of electron spectrometer is provided by retarding field analysers, which consist of a set of concentric hemispherical grids \cite{Palmberg}. While this technique is significantly easier to apply, the achieved energy resolution is typically in the range of a few 100~meV. The expected low signal-to-background ratio of this technique will again make short ion-beam pulsing an important tool.\\
Independent of the applied technique for electron spectroscopy, charge exchange is required for the thorium ions in order to trigger the IC decay. In the simplest approach, this charge exchange is achieved by deposition of the thorium ions on a surface. This technique, however, is expected to influence the energy of the IC electrons as the work function of the surface material has to be considered \cite{Hotop}. In case of CsI, as being the coating material of the MCP used for all presented detections, the work function is 6.2~eV and thus close to the first ionisation potential of thorium \cite{Poole}. For this reason no significant influence on the reported energy of the IC electrons is expected. For a precise energy determination, however, a careful investigation of surface influences is required. This must include the collection of the thorium ions on different surface materials with different work functions.\\
An alternative to the collection on a surface could be provided by collision with an atom beam. By crossing the thorium ion beam with a beam of, for example, caesium atoms, charge exchange will trigger the isomeric decay and could lead to an improved energy determination as no surface influences have to be considered \cite{Yamakita}.

\end{small}

\end{multicols*}
\renewcommand{\figurename}{Extended Data Figure}
\renewcommand{\tablename}{Extended Data Table}
\setcounter{figure}{0}

\begin{figure}[H]
\begin{center}
\includegraphics[width=18.3cm]{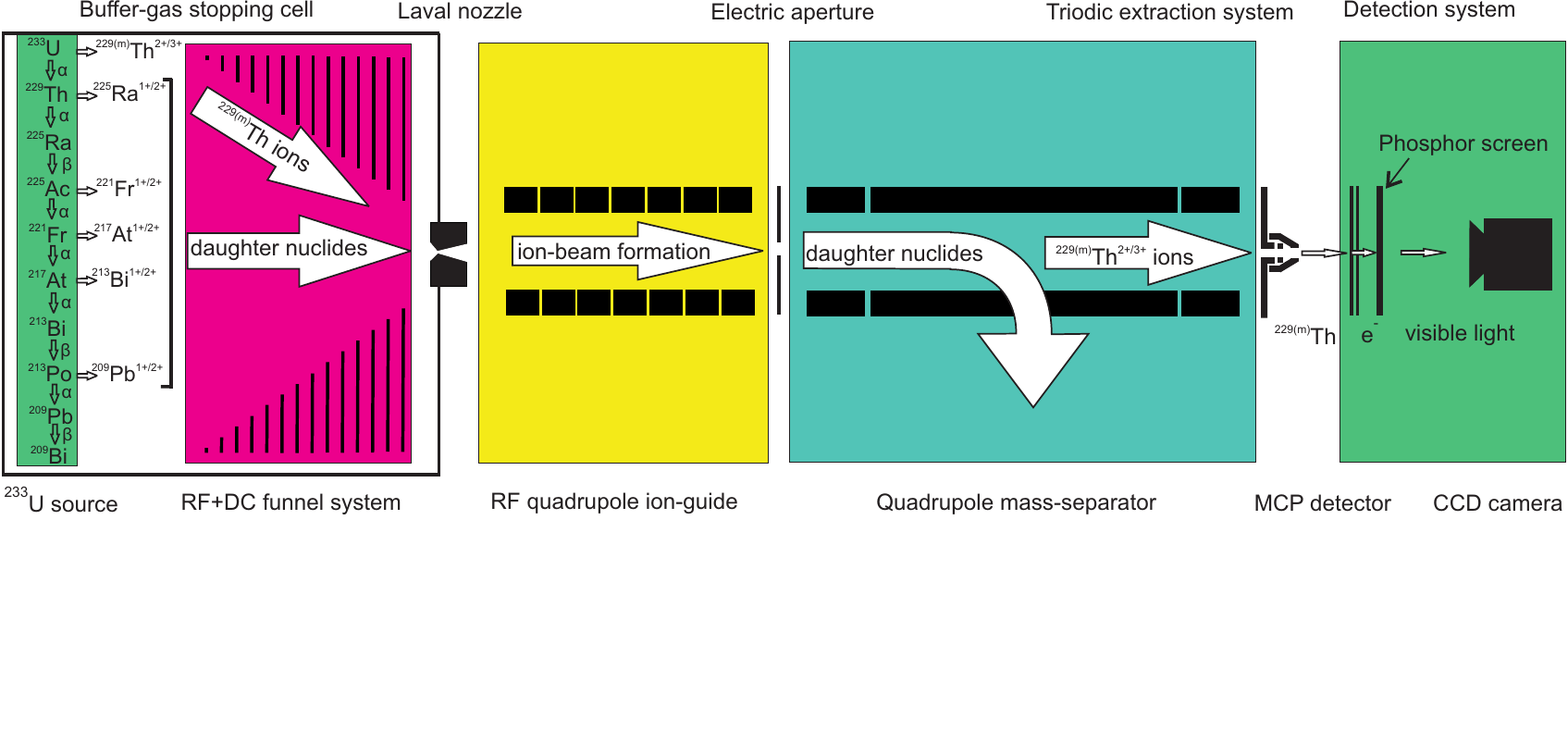}
\vspace{-2.5cm}
\caption{{\bf Schematic of the experimental process.} Daughter nuclides of the $^{233}$U decay chain leave the $^{233}$U source due to the kinetic recoil energy transferred to the nucleus during the $\alpha$ decay. Only those nuclides produced by $\alpha$ decay have enough kinetic recoil energy for efficiently leaving the $^{233}$U source material. The maximum layer thickness, through which recoiling nuclei can pass, is a few ten nanometeres. The $\alpha$-recoil nuclei are thermalised with helium and extracted from the stopping cell. The process of electron capture during thermalisation leads to the formation of ions in the $1+$, $2+$ or $3+$ charge states. Subsequently, an ion beam is formed and purified with a quadrupole mass-separator such that only $^{229}$Th remains. The thorium ions are collected in soft landing on the surface of a micro-channel-plate (MCP) detector and the isomeric decay is detected.}
 \label{scheme}
 \end{center}
\end{figure}

\clearpage

\begin{figure}[H]
 \begin{center}
 \includegraphics[width=18.3cm]{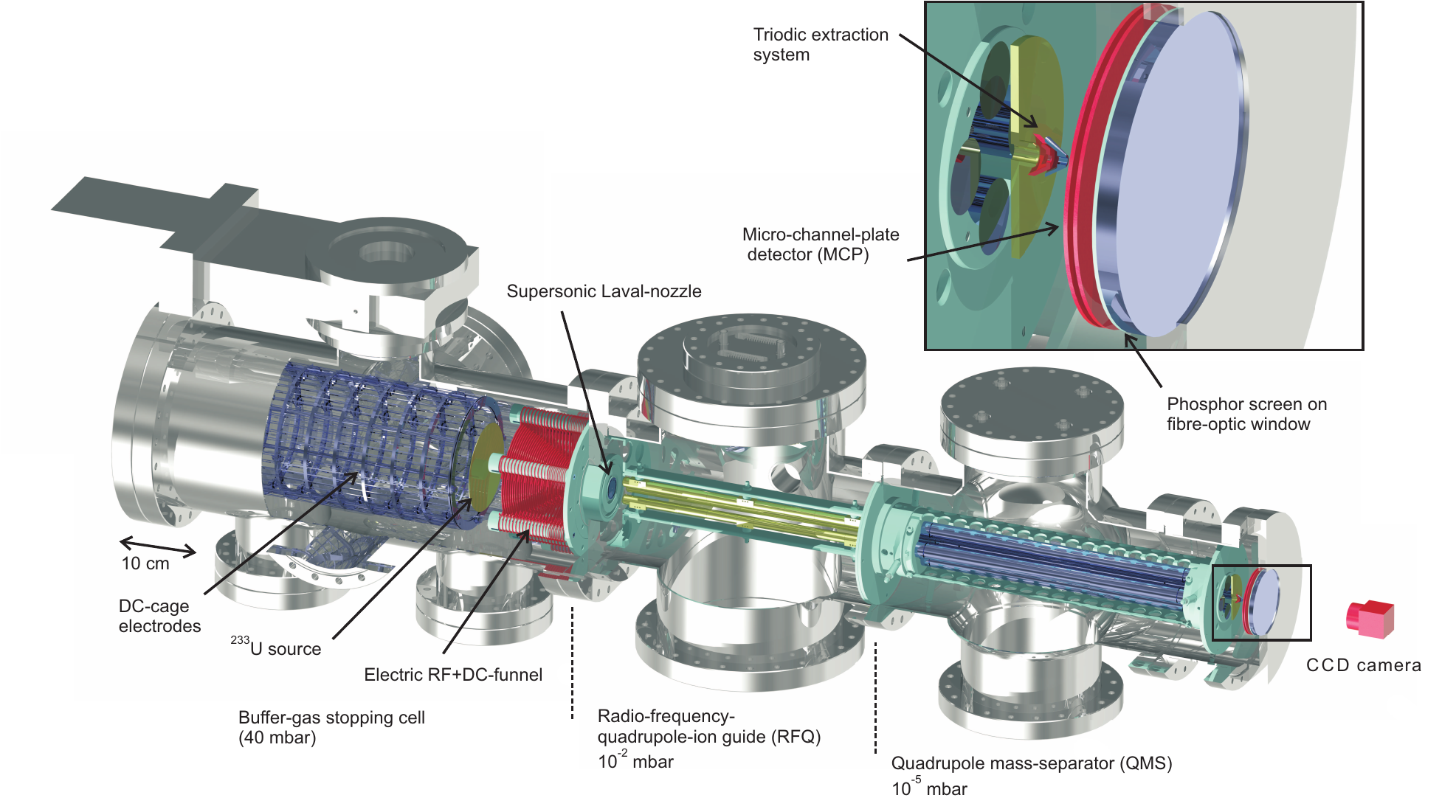}
  \caption{ {\bf Overview of the experimental setup.} The buffer-gas stopping cell houses the $^{233}$U source, which is mounted onto the front end of a DC-cage-electrode system \cite{Neumayr2}. The $^{229}$Th $\alpha$-recoil ions emitted from the source are stopped in the buffer-gas stopping cell filled with 40~mbar helium. These ions are then guided by an electric RF+DC funnel system towards the exit of the stopping cell formed by a supersonic Laval-nozzle, injecting them into a radio-frequency quadrupole (RFQ) ion guide, where an ion beam is formed by phase-space cooling due to the remaining helium pressure of $10^{-2}$~mbar. Following the RFQ, the ion beam is purified after a mass-to-charge separation with a quadrupole mass-separator (QMS). Behind the QMS a micro-channel-plate (MCP) allows for the detection of the low-energy internal conversion (IC) electrons emitted in the $^{229}$Th isomeric decay.}
 \label{buffergascell}
 \end{center}
\end{figure}

\clearpage

\begin{figure}[H]
 \begin{center}
 \includegraphics[width=18.3cm]{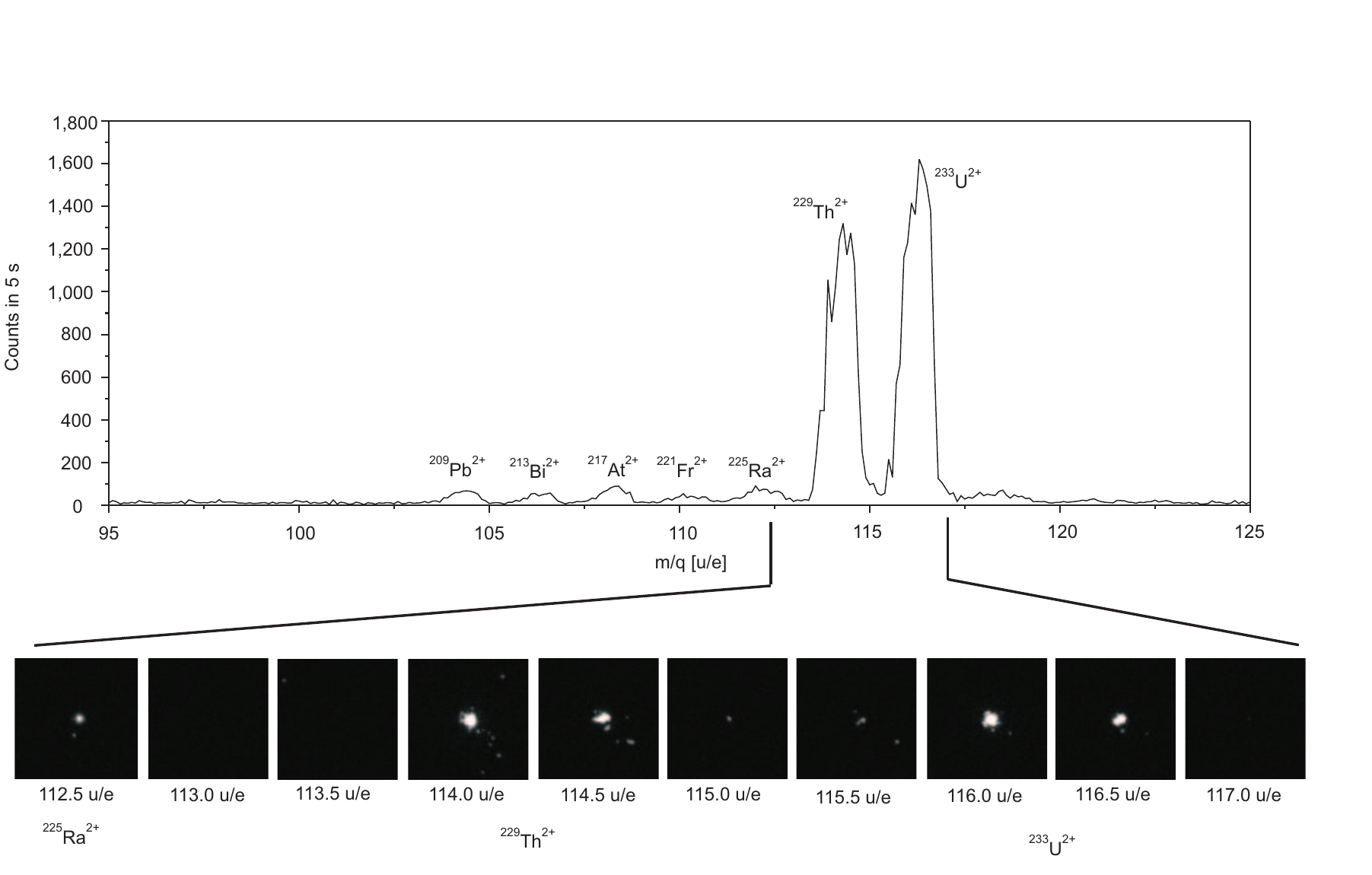}
  \caption{{\bf Intensity profile measurements.} Upper panel: Mass spectrum in the range of the $2+$ ion species as performed with the chemically unpurified $^{233}$U source 1 and an MCP detector (Hamamatsu, type F2223) operated in single-ion counting mode. Lower panel: Ion impact profile measurement ($-900$~V MCP surface voltage, 1~s exposure time) performed with $^{233}$U source 1 and an MCP detector allowing for spatially resolved read-out (Beam Imaging Solutions: BOS-75-FO). The $^{229}$Th and $^{233}$U mass peaks can clearly be separated.}
 \label{lightintensity}
 \end{center}
\end{figure}

\clearpage

\begin{figure}[H]
 \begin{center}
 \includegraphics[width=18.3cm]{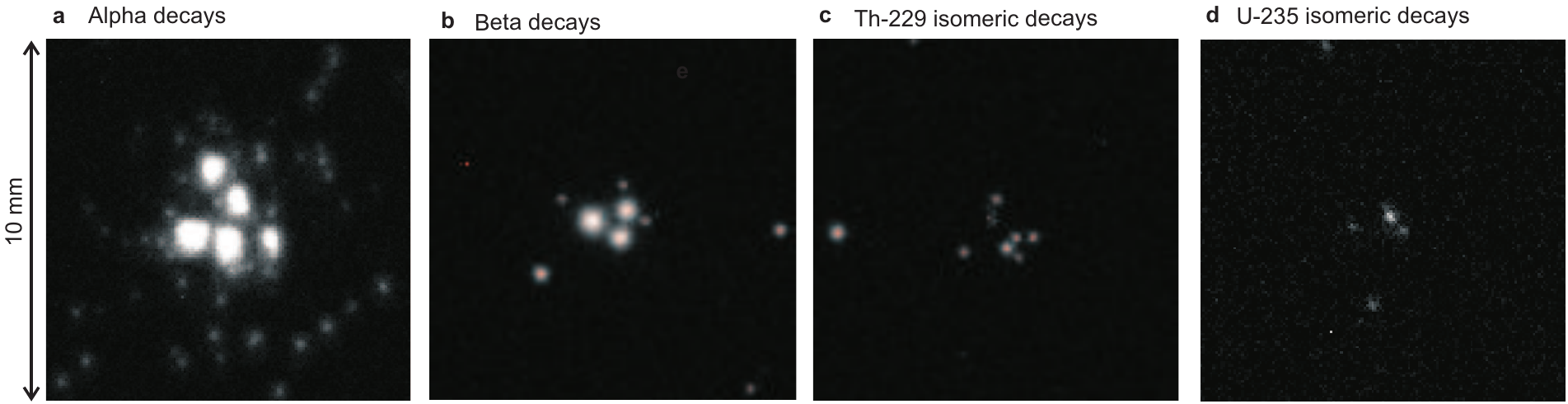}
  \caption{{\bf Different classes of decay events as observed during ion accumulation on the MCP surface.} In order to suppress any ion-impact signal, soft landing of the ions is guaranteed at $-25$~V MCP surface voltage. Single frames of 4~s exposure time are shown. The MCP detector used (Beam Imaging Solutions: BOS-75-FO) allows for spatially resolved image read-out. The extracted ion species is chosen by mass-to-charge separation with the help of the QMS. a: Alpha decays originating from $^{221}$Fr. b: Beta decays originating from $^{209}$Pb. c: Isomeric decay of $^{229}$Th. d: Isomeric decay of $^{235}$U. In the shown frames all ions were extracted in the $2+$ charge state from the chemically unpurified $^{233}$U source 1.}
 \label{MCP_events}
 \end{center}
\end{figure}

\newpage

\begin{figure}[H]
 \begin{center}
 \includegraphics[width=18.3cm]{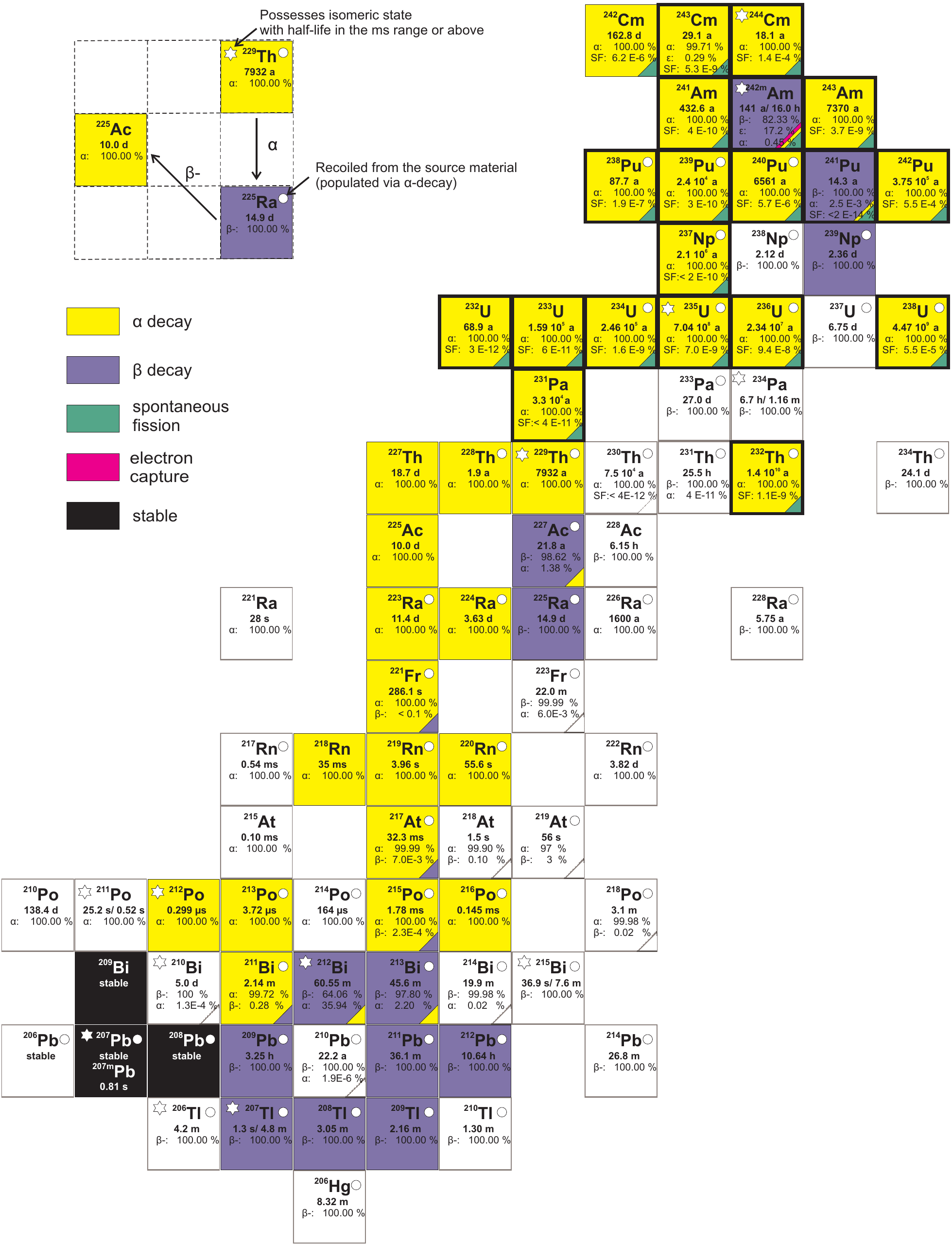}
  \caption{{\bf Chart of nuclides as being potentially contained in the source material.} The chart includes all elements from curium (Cm, $Z = 96$) to mercury (Hg, $Z = 80$). All nuclides drawn are taken into consideration for the exclusion of a potential nuclear background. For completeness, all potentially populated nuclides are shown, even if their activity can be assumed to play a negligible role due a small branching ratio or a long half-life of the mother nuclide. These nuclides are shown without color. Nuclides, which can potentially recoil from the source as populated via $\alpha$ decay, are assigned with a white circle. Nuclides, which possess one or more isomeric states, carry a white star. A complete list of potentially contributing excited isomers is given in Extended Data Table~3.}
 \label{decay_chain}
 \end{center}
\end{figure}

\newpage

\begin{figure}[H]
 \begin{center}
 \includegraphics[width=18.3cm]{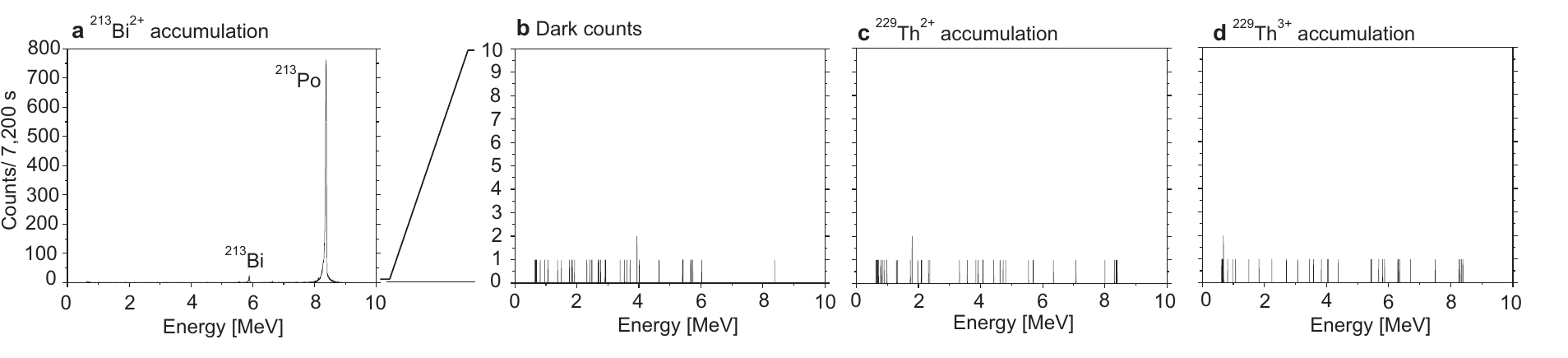}
  \caption{{\bf $\alpha$-energy spectra of different Si-detector-based measurements, each accumulated for 7,200~s.} A silicon charged particle detector (Ametek, BU-014-150-100) is used for detection. The extracted ion species is chosen by mass-to-charge separation by the QMS. The accumulated counts are shown for extraction from the chemically unpurified $^{233}$U source 1 for a: $^{213}$Bi$^{2+}$, b: no extraction, c: $^{229}$Th$^{2+}$, d: $^{229}$Th$^{3+}$. No signal above the background is detected for $^{229}$Th in the $2+$ and $3+$ charge states. This clearly excludes any $\alpha$ decay as signal origin.}
 \label{simeas}
 \end{center}
\end{figure}

\clearpage

\newcolumntype{C}[1]{>{\centering\arraybackslash}m{#1}}
\begin{table}[H]
\begin{center}
\begin{footnotesize}
\caption{{\bf Matrix of potential background contributions and the different ways of their exclusion.} Column 1 lists the measurement numbers as given in the text, while in column 2 the corresponding measurement types are detailed. From 3 to 6 each column corresponds to one type of potential background contribution. A cross indicates its exclusion by the measurement given in the corresponding row. Most of the potential background contributions could be excluded in multiple ways. A: Background from ionic kinetic energy or energy carried in form of charge state of the impinging ion. B: Background originating from the setup components ($^{233}$U source, ion transport system, detection system). C: Background from the thorium atomic shell. D: Background from activity other than $^{229}$Th.}
\vspace{0.5cm}
\begin{tabular}{|c|c|C{0.6cm}|C{0.6cm}|C{0.6cm}|C{0.6cm}|}
\hline
\multirow{2}{*}{No.} & \multirow{2}{*}{Way of background exclusion} & \multicolumn{4}{c|}{Type of background}\\
\cline{3-6}
& & A & B & C & D \\
\hline
1 & \parbox[0pt][3.5em][c]{4.7cm}{Signal comparison between~$^{229}$Th$^{2+}$ and $^{233}$U$^{2+}$} & x & x &  &\\
\hline
2 & \parbox[0pt][3.5em][c]{4.7cm}{Comparative $^{229}$Th$^{2+}$ and $^{233}$U$^{2+}$ signal behaviour as a function of MCP surface voltage} & x & x &  &\\
\hline
3 & \parbox[0pt][3.5em][c]{4.7cm}{Signal comparison between $^{229}$Th and $^{230}$Th} & x & x & x & \\
\hline
4 & \parbox[0pt][3.5em][c]{4.7cm}{Signal comparison between~$^{229}$Th$^{2+}$ and $^{229}$Th$^{3+}$} &  &  & & x\\
\hline
5 & \parbox[0pt][3.5em][c]{4.7cm}{Signal comparison between $^{229}$Th originating from chemically purified and unpurified $^{233}$U sources} &  &  &  & x\\
\hline
6 & \parbox[0pt][3.5em][c]{4.7cm}{Exclusion based on signal appearance and the $^{229\mathrm{m}}$Th half-life limit} &  &  & & x\\
\hline
7 & \parbox[0pt][3.5em][c]{4.7cm}{Search for $\alpha$ and $\beta$ decays using Si and LN$_2$ cooled Si(Li) detectors} &  &  &  & x\\
\hline
\end{tabular}
\label{thorium_extdat_3}
\end{footnotesize}
\end{center}
\end{table}

\newpage

\begin{table}[H]
\begin{center}
\begin{footnotesize}
\caption{{\bf List of ionisation energies \cite{ionize} of the first 3 charge states of elements potentially contained in the $^{233}$U source material.} From radium downwards all elements reveal 3$^{\mathrm{rd}}$ ionisation potentials which are above the 1$^{\mathrm{st}}$ ionisation potential of helium (E$_{\mathrm{ion}}$=24.6~eV). Besides the mass-to-charge separation, this feature is also exploited to remove short-lived nuclides from the $^{229}$Th$^{3+}$ ion beam, as only elements with a 3$^{\mathrm{rd}}$ ionisation potential below 24.6~eV can be extracted from the buffer-gas stopping cell to a significant amount in the $3+$ charge state. For other elements the $3+$ charge state is reduced to the $2+$ charge state during collisions with the helium buffer-gas.}
\vspace{0.5cm}
\begin{tabular}{l c D{.}{.}{-1} D{.}{.}{-1} D{.}{.}{-1}}
\hline
Element & Atomic no. & \multicolumn{1}{c}{\text{1+ [eV]}} & \multicolumn{1}{c}{\text{2+ [eV]}} & \multicolumn{1}{c}{\text{3+ [eV]}} \\
\hline

Curium & 96 &5.99 & 12.4 & 20.1 \\

Americium & 95 & 5.97 & 11.7 & 21.7 \\

Plutonium & 94 & 6.03 & 11.5 & 21.1 \\

Neptunium & 93 & 6.27 & 11.5 & 19.7 \\

Uranium & 92 & 6.19 & 11.6 & 19.8 \\

Protactinium & 91 &  5.89 & 11.9 & 18.6 \\

Thorium & 90 & 6.31 & 11.9 & 18.3 \\

Actinium & 89 & 5.38 & 11.8 & 17.4 \\

Radium & 88 & 5.28 & 10.1 & 31.0 \\

Francium & 87 & 4.07 & 22.4 & 33.5 \\

Radon & 86 & 10.75 & 21.4 & 29.4 \\

Astatine & 85 & 9.32 & 17.9 & 26.6 \\

Polonium & 84 & 8.41 & 19.3 & 27.3 \\

Bismuth & 83 & 7.29 & 16.7 & 25.6 \\

Lead & 82 & 7.42 & 15.0 & 31.9 \\

Thallium & 81 & 6.11 & 20.4 & 29.9 \\

Mercury & 80 & 10.44 & 18.7 & 34.5 \\
\hline
\end{tabular}
\label{thorium_extdat_5}
\end{footnotesize}
\end{center}
\end{table}

\newpage

\begin{table}[H]
\begin{center}
\begin{footnotesize}
\caption{{\bf List of known isomeric states of nuclides potentially contained in the $^{233}$U source material.} The isomeric excitation energies, half-lives, decay channels and population branching ratios are listed \cite{gov}. SF is short for spontaneous fission and IT for internal transition, which includes both ways of deexcitation: by photon emission or by internal conversion.}
\vspace{0.5cm}
\begin{tabular}{l c c c c }
\hline
Isomer & Excitation energy & Half-life & Decay channel & Population\\
\hline
$^{244\mathrm{m}}$Cm & 1.04 MeV & 34 ms & IT: 100.00 \% & not populated \\

$^{242\mathrm{m1}}$Am & 48.6 keV & 141 a & IT: 99.55 \%, $\alpha$: 0.45 \% & 100 \% populated \\
$^{242\mathrm{m2}}$Am & 2.20 MeV & 14.0 ms & SF: 100 \%, $\alpha:\ \le 5.0\cdot 10^{-3}$ \%, IT& not populated\\

$^{235\mathrm{m}}$U & 76 eV & 26 min & IT: 100 \% & 70 \% from $^{239}$Pu\\

$^{234\mathrm{m}}$Pa & 73.9 keV & 1.16 min & $\beta-$: 99.84 \%, IT: 0.16 \% & 78 \% from $^{234}$Th\\

$^{229\mathrm{m}}$Th & $\sim$7.8 eV & unknown & unknown & 2 \% from $^{233}$U\\

$^{212\mathrm{m}}$Po & 2.922 MeV & 45.1 s & $\alpha$: 99.93 \%, IT: 0.07 \% & not populated\\

$^{211\mathrm{m}}$Po & 1.462 MeV & 25.2 s & $\alpha$: 99.98 \%, IT: 0.02 \% & not populated\\

$^{215\mathrm{m}}$Bi & 1.348 MeV & 36.9 s & IT: 76.2 \%, $\beta-$: 23.8 \% & not populated\\

$^{212\mathrm{m1}}$Bi & 0.250 MeV & 25.0 min & $\alpha$: 67.0 \%, $\beta-$: 33.0 \% & not populated\\
$^{212\mathrm{m2}}$Bi & 1.91 MeV  & 7.0 min & $\beta-$: 100 \% & not populated\\

$^{210\mathrm{m}}$Bi & 0.271 MeV & $3.04 \cdot 10^6$ a & $\alpha$: 100 \% & not populated\\

$^{207\mathrm{m}}$Pb & 1.633 MeV & 0.806 s & IT: 100 \% & $8.1 \cdot 10^{-4}$ \% from $^{211}$Po \\

$^{207\mathrm{m}}$Tl & 1.348 MeV & 1.33 s & IT: 100 \% & $9 \cdot 10^{-4}$ \% from $^{211}$Bi\\

$^{206\mathrm{m}}$Tl & 2.643 MeV & 3.74 min & IT: 100 \% & not populated\\
\hline 
\end{tabular}
\label{thorium_extdat_8}
\end{footnotesize}
\end{center}
\end{table}

\end{document}